\documentclass{ws-procs9x6}

\newcommand{\eqnlessref}[1]{(\ref{#1})}

\def\Det{{\rm Det}}

\def\scrD{{\mathcal D}}
\def\lsim{\mathrel{\lower0.3em\hbox{$\stackrel{\textstyle <}{\sim}$}}}
\def\gsim{\mathrel{\lower0.3em\hbox{$\stackrel{\textstyle >}{\sim}$}}}
\def\scrK{{\mathcal K}}
\def\negspace{\kern -0.4em}

\def\dvec{\raise 0.3 em \hbox{$^\leftrightarrow$} \kern -0.77 em}

\def\omegahat{\hat%
	{\setbox0=\hbox{$\omega$}%
		\kern-.025em\copy0\kern-\wd0
		\kern.05em\copy0\kern-\wd0
		\kern-.025em\raise.0433em\box0}}

\begin{document}
\title{From QCD to Dual Superconductivity to Effective String Theory}

\author{M. Baker}
\address{Department of Physics, University of Washington, P.O. Box 351560, Seattle, WA 98195, USA\\}

\maketitle\abstracts{
    We show how an effective field theory of long distance QCD,
    describing a dual superconductor,
    can be expressed as an effective string theory of superconducting
    vortices. We use the semiclassical expansion of this effective
    string theory about a classical rotating string solution
    in any spacetime dimension $D$ to obtain the semiclassical meson 
    energy spectrum. We argue that the experimental data on Regge 
    trajectories along with numerical simulations of the heavy quark 
    potentials provide good evidence for an effective string 
    description of long distance QCD.}


\section{From QCD to Dual Superconductivity}

In the dual superconductor mechanism of 
confinement~\cite{Nambu,Mandelstam,tHooft} a dual Meissner
effect confines color electric flux 
to narrow tubes~\cite{Nielsen+Olesen} connecting a quark-antiquark pair. 
In the confined
phase, monopole fields $\phi$ condense to a
value $\phi_0$, and dual potentials $C_\mu$
acquire a mass $M = g\phi_0$ via a dual Higgs mechanism.
The dual coupling constant is $g=2\pi/e$, where $e$ is
the Yang--Mills coupling constant. 
Quarks couple to dual potentials via a Dirac string connecting
the quark-antiquark pair along a line L, the ends of which 
are sources and sinks of color electric flux. The color field of the 
pair destroys the dual Meissner effect near L so that $\phi$
vanishes on L. At distances transverse to L greater than $1/M$
the monopole field returns to its bulk value $\phi_0$, so that 
the color field is confined to a tube of radius $a = 1/M$ surrounding 
the line L.
As a result, for quark-antiquark separations $R$ greater than $a$, 
a linear potential develops that confines the quarks in hadrons.
The string tension $\sigma \sim {\phi_0}^2$ , so that $M \sim \sqrt {\sigma/\alpha_s}$.
(The running coupling $\alpha_s$ is evaluated at a scale of order $M$.)

Recent lattice calculations~\cite{Forcrand} and general arguments based on the work of 't Hooft~\cite{tHooft2} show that the confined phase
of non-Abelian gauge
theory is characterized by a dual order parameter (monopole condensate), which
vanishes in regions of space where electric color flux can penetrate
and "dual supercondutivity" 
is destroyed. This provides a basis in QCD for a generic dual 
superconducting effective field theory of long distance Yang Mills 
theory in which the dual gluon mass $M$ serves 
as the ultraviolet cutoff.

To obtain an effective string theory of long distance QCD we do 
not need a specific form for the action 
$S[C_\mu,\phi]$ of the effective dual field theory.
This theory
must have classical $Z_N$ electric vortex 
solutions, and it cannot
have massless particles.
 For $SU(N)$ Yang Mills theory this can be done
by coupling dual non-Abelian potentials $C_\mu$ to $Z_N$
invariant dual Higgs fields $\phi$ . 
(For example there could be Higgs fields in the adjoint representation 
of the gauge group.)
The long distance
 effective dual theory will then have the same symmetries 
as $SU(N)$ Yang Mills theory and
the dual $SU(N)$ gauge symmetry can be "spontaneously broken"
so that the gauge bosons $C_\mu$ all acquire a mass, and 
there are $Z_N$ electric flux tube excitations.~\cite{Baker+Ball+Zachariasen:1991}

In the classical approximation to the dual theory the axis of the flux 
tube is a straight line between the quark and the antiquark.
The contribution of flux tube fluctuations to the heavy quark potential
is determined by the path integral over all field
configurations $C_\mu,\phi$, for which the monopole fields $\phi$ 
vanish on some line $L$ connecting the quark-antiquark pair.~\cite{Nora}
The fluctuating vortex line $L$ sweeps out a fluctuating 
spacetime surface $\tilde x^\mu$, whose boundary is the loop $\Gamma$
formed by the worldlines of the moving pair.  These surfaces  
$\tilde x^\mu$ determine the location of the vortices, where dual
superconductivity is destroyed and electric color flux can penetrate.

The Wilson loop $W[\Gamma]$ of Yang Mills theory determining the quark-antiquark interaction
is the partition function of the dual theory in the vortex sector.~\cite
{Nora}
\begin{equation}
W[\Gamma] = \int \scrD C_\mu \scrD\phi \scrD\phi^* e^{iS[C_\mu, \phi]}
 \,.
\label{Wilson loop def}
\end{equation}
The path integral \eqnlessref{Wilson loop def} goes over all field
configurations for which the monopole field  $\phi(x)$
vanishes on some sheet $\tilde x^\mu$ bounded by the loop
$\Gamma$.

\section{From Dual Superconductivity to Effective String Theory}

We transform the field theory partition function
\eqnlessref{Wilson loop def} 
to a path integral over the vortex sheets $\tilde x^\mu$, so
that $W[\Gamma]$ takes the form of the  partition function
of an effective string theory of these vortices. We do this
in two stages:

\begin{enumerate}
\item We integrate over all field configurations $C_\mu,\phi$,
containing a vortex located on a particular surface
$\tilde x^\mu$, where $\phi(\tilde x^\mu) = 0$. 
This integration determines the action $S_{{\rm eff}}[\tilde x^\mu]$
of the effective string theory.
\item We integrate over all
vortex sheets $\tilde x^\mu(\xi)$, $\xi=\xi^a$, $a=1\,,2$. 
This integration goes over the amplitudes $f^1(\xi)$ and $f^2(\xi)$
of the two transverse vortex fluctuations
in a particular parameterization 
of the world 
sheet,  $\tilde x^\mu(\xi) = x^\mu(f^1(\xi), f^2(\xi), \xi )$ , and
gives~\cite{Baker+Steinke2}
\begin{equation}
W[\Gamma] = \int \scrD f^1 \scrD f^2 \Delta_{FP} e^{iS_{{\rm eff}}[\tilde x^\mu]} \,,
\label{Wilson loop eff}
\end{equation}
where
\begin{equation}
\Delta_{FP} = \Det\left[ \frac{\epsilon_{\mu\nu\alpha\beta}}{\sqrt{-g}}
\frac{\partial x^\mu}{\partial f^1} \frac{\partial x^\nu}{\partial f^2}
\frac{\partial \tilde x^\alpha}{\partial \xi^1}
\frac{\partial \tilde x^\beta}{\partial \xi^2}
\right]
\end{equation}
and $\sqrt{-g}$ is the square root
of the determinant of the induced metric 
\begin{equation}
g_{ab} = \frac{\partial \tilde x^\mu}{\partial \xi^a}
\frac{\partial \tilde x_\mu}{\partial \xi^b} \,.
\end{equation}
\end{enumerate}
The partition function \eqnlessref{Wilson loop def}
of an effective field theory has been expressed as the
partition function \eqnlessref{Wilson loop eff} of an effective
string theory.
The presence of the determinant $\Delta_{FP}$ in 
\eqnlessref{Wilson loop eff}  
makes the
path integral invariant under reparameterizations $\tilde x^\mu(\xi)
\to \tilde x^\mu(\xi(\xi^\prime))$ of the vortex worldsheet. 

The parameterization invariant measure in the path 
integral \eqnlessref{Wilson loop eff} is universal and is
independent of the explicit form of the underlying field theory.
On the other hand, the action $S_{{\rm eff}}[\tilde x^\mu]$ 
of the effective string theory is 
not universal, and depends upon parameters in the action 
$S[C_\mu,\phi]$ 
of the effective field theory describing the dual superconductor.
However, for wavelengths $\lambda$ of the string fluctuations greater 
than the flux tube radius $a=1/M$, which are those included 
in \eqnlessref{Wilson loop eff}, the action 
$S_{{\rm eff}}[\tilde x^\mu]$ can be expanded in powers of
the extrinsic curvature tensor $\scrK^A_{ab}$ of the sheet
$\tilde x^\mu$,
\begin{equation}
S_{{\rm eff}}[\tilde x^\mu] = - \int d^4x \sqrt{-g} \left[ \sigma
+ \beta \left(\scrK^A_{ab}\right)^2 + ... \right] \,.
\label{curvature expansion}
\end{equation}
The extrinsic curvature tensor is
\begin{equation}
\scrK^A_{ab} = n^A_\mu(\xi) \frac{\partial^2 \tilde x^\mu}
{\partial\xi^a \partial\xi^b} \,,
\end{equation}
where $n^A_\mu(\xi)$, $A = 1,2$ are vectors normal to
the worldsheet at the point $\tilde x^\mu(\xi)$.
The values of the coefficients in this expansion (
the string tension $\sigma$, the rigidity $\beta$, ...) are
determined by the parameters of the underlying effective field theory,
(For example in a dual superconductor on the border between type I
and type II the rigidity vanishes.)~\cite{Baker+Steinke3}
If these coefficients are taken as parameters, the specific form of the
dual field theory does not enter explicitly in the effective string 
theory of dual superconductivity \eqnlessref{Wilson loop eff}.
The expansion parameter in \eqnlessref{curvature expansion}
is the ratio $(a/\lambda)^2$ of the square of the 
flux tube radius to the square of 
the wave length of the string fluctuations, and the leading term in this
expansion is the Nambu--Goto action.

\section{The Heavy Quark Potentials}

Expanding the action
\eqnlessref{curvature expansion} in small fluctuations about a straight 
string connecting two static quarks puts the partition function
\eqnlessref{Wilson loop eff} in the form used by L{\"u}scher, Symanzik
and Weisz~\cite{Luscher1,Luscher2,Luscher3} to calculate
the contribution of string fluctuations to the static heavy quark potentials. The
leading long distance expression for the energy levels $E_n(R)$
of a fixed string of length $R$, vibrating in D dimensional spacetime
is~\cite{Luscher3}
\begin{equation}
E_n(R) = {\sigma}{R} + (-\frac{D-2}{24} + n)\frac{\pi}{R} \,.
\label{static energy}
\end{equation}
Corrections to \eqnlessref{static energy} arise from the higher order 
terms in $S_{{\rm eff}}[\tilde x^\mu]$ , which give contributions 
proportional to powers of $(a/R)^2$.
For $n=0$, \eqnlessref{static energy} reduces to the L{\"u}scher 
ground state heavy quark potential.

Recent numerical simulations~\cite{Luscher3,Sommer} provide
striking confirmation of the L{\"u}scher potential for 
quark-antiquark separations greater $0.5$ fm. At this distance 
and (with $D=4$) the ratio
$\pi/(12 \sigma R^2)$ of the leading semiclassical correction in eq.
\eqnlessref{static energy} (with $n=0$)
to the classical term $\sigma R$ is already small ($\sim 0.2$). Thus the 
string behavior of the static potential sets in at a distance where
the semiclassical expansion parameter is small.  

The excited potentials $E_n(R)$, $n>0$, involve wave 
lengths of order $R/n$. When this wave length is of order of the flux 
tube radius, higher order terms in the effective action should become 
important and modify the string behavior of the excited potentials.
Therefore as $n$ increases, the distance $R$ for which 
\eqnlessref{static energy} should be applicable becomes
larger. This expectation 
is in qualitative agreement with the recent lattice measurements of
Juge, Kuti, and Morningstar~\cite{Kuti}
of the excited energy levels. 
They find that, although the excited potentials 
do not have string behavior at $R=0.5$ fm,
there is, for $R \approx 2$ fm, 
a rapid rearrangement of
excited levels towards the string ordering 
\eqnlessref{static energy}. There are questions to be 
clarified, but these measurements of excited  heavy quark 
potentials provide further evidence for an effective string theory 
description of long distance QCD.

\section{Meson Regge Trajectories}

\begin {figure}
    \begin{center}
	\null \hfill \epsfbox{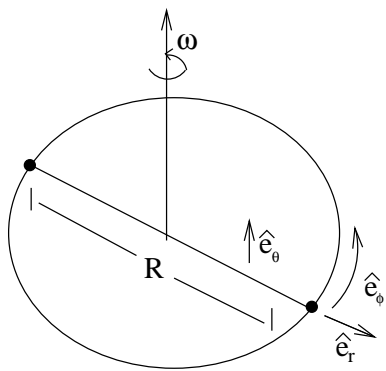} \hfill \null
    \end{center}
    \caption{The string coordinate system}
    \label{rot string fig}
\end {figure}

We can also use the effective string theory to calculate the Regge
trajectories of light mesons by attaching massless
scalar quarks to the ends of a rotating string.~\cite{ron}
(We treat neither chiral symmetry breaking nor string breaking.)
Consider a quark-antiquark pair rotating with uniform angular velocity $\omega$
and separated by a distance $R$.
(See Fig.~\ref{rot string fig}).
Calculating the classical energy and angular momentum of a straight 
rotating 
string gives the usual linear Regge trajectories.
To obtain the contribution
of string fluctuations to these classical trajectories
we expand the action $S_{{\rm eff}}[\tilde x^\mu]$
in small fluctuations $ f^i $
about a classical rotating straight string solution.
In D spacetime dimensions there are
$D-3$ fluctuations perpendicular to the plane of rotation and 
there is 1 fluctuation in the plane of rotation.

The semiclassical calculation of the path
integral \eqnlessref{Wilson loop eff} 
around the classical rotating straight string solution 
with massless quarks
contains an infrared divergence, so we introduce a quark mass 
as an infrared cutoff which can be set equal to zero only after renormalization.  
Furthermore, since the classical world sheet is no longer flat,
in addition to the quadratic and linear ultraviolet divergences
present in the calculation of the energy of the string fluctuations
about a static string,
the path integral \eqnlessref{Wilson loop eff} now
contains a term which diverges logarithmically in the ultraviolet 
cutoff $M$. 
This divergence can be absorbed into a renormalization of a term in the
boundary action called
the geodesic curvature \cite{Alvarez:1983}. 

The path integral \eqnlessref{Wilson loop eff}
can then  be evaluated
for any values of the masses of the quarks on the end of the string and
gives the energy levels of the interior of the rotating string for a 
prescribed motion of its ends. The frequency $\omega$ determines the
interquark distance $R$ . These energy levels reduce to the heavy
quark potentials \eqnlessref{static energy} if $\omega$ is set equal 
to zero with $R$ held fixed. 
For massless rotating quarks, these levels can be obtained from
\eqnlessref{static energy} by replacing the length $R$ 
of the string by its proper length $ \pi/\omega $ ,
and by adding the term $ -\omega /2 $, 
which enters because the classical background is not flat.
This gives~\cite{ron} 

\begin{equation}
E_n(\omega) = \frac{\pi\sigma}{\omega} + ( -\frac{D-2}{24} 
+ n )\omega  -\frac{\omega}{2} \,.
\label{hybrid energy ll}
\end{equation}

To obtain the meson energy levels we must also quantize the motion
of the quarks on the ends of the string.~\cite{ron} For massless quarks,
accounting for these boundary fluctuations effectively changes a Dirichlet
boundary condition into a Neumann boundary condition.
The meson energy levels are given by the same expression
\eqnlessref{hybrid energy ll} and the meson angular momentum $J$ is
quantized,
\begin{equation}
J = 
 l + \frac{1}{2} , \kern 1 in l = 0, 1, 2,... \,.
\label{J quant}
\end{equation}

 The value of $\omega$ is given as a function of $J$ through the
classical relation $\omega = \sqrt{\pi\sigma/2J}$. Squaring
both sides of \eqnlessref{hybrid energy ll} and using
the WKB quantization condition \eqnlessref{J quant} yield
the sequence of linear Regge trajectories,
\begin{equation}
E_n^2(l)  = 2\pi\sigma\left(l - \frac{D-2}{24} +n \right) ,
 \kern 1 in n =0, 1, 2, ... \,.
\label{E^2 l}
\end{equation}
Corrections to 
\eqnlessref{E^2 l} of order $ n^2/l $ come from the square of the 
term linear in $ \omega $
in \eqnlessref{hybrid energy ll} and from higher order terms 
in the semiclassical expansion.

Eq. \eqnlessref{E^2 l}, valid in any spacetime dimension , but applicable
only for large angular momentum $l$ gives, for $n=0$, the
contribution of string fluctuations to the leading Regge trajectory.
The ratio $(D-2)/24l$ of the leading semiclassical correction
to the classical term is already small for $l=1$. 
This could provide an explanation for the approximate experimental
linearity of leading light meson Regge trajectories for $l$ 
of the order $1$, in analogy with the expectation from 
\eqnlessref{static energy} that the string behavior of the static
potential should set in at a distance $R \sim 0.5 $ fm.

The energies \eqnlessref{E^2 l} (with $n>0$) of the excited states
of the rotating string give 
rise to daughter Regge trajectories, and the calculation
is applicable when $l$ is much greater than $n^2$ .
According to this picture, linear daughter trajectories should be 
expected only for values of $l$ much greater than 1. This corresponds
to the expected linear behavior of the excited static potentials \eqnlessref{static energy} only at values of $R$ much greater than $0.5$ fm.


\section{Summary and Conclusions}

\begin{enumerate}
\item We have found a path, QCD $\rightarrow$  Effective Field Theory of
Dual Superconductivty $\rightarrow$  Effective String Theory, 
 from QCD to effective string theory,
which provides a concrete picture of the QCD string.


\item The derivation of the effective string theory
made no use of the details of the effective dual field theory
from which it was obtained, but the arguments~\cite{Forcrand,tHooft2}
leading from QCD to an effective dual field theory 
description of long distance QCD need to be developed further.
\item The effective string theory provides an
understanding of the values of the distances and of
the angular momentum at which string behavior of 
physical quantities sets in.

\end{enumerate}

\section*{Acknowledgements}

I would like to thank Ph. de Forcrand and L. Yaffe for numerous enlightening conversations,
M. L{\"u}scher and P. Weisz for helpful comments, and the organizers 
of this conference for making it so stimulating and pleasant.



\end{document}